\documentclass[pra,showkeys]{revtex4}

\usepackage{bm}
\usepackage{graphicx}
\usepackage{amssymb}
\usepackage{amsmath}
\usepackage{amsthm}
\usepackage{multirow}
\usepackage{mathrsfs}
\DeclareSymbolFont{AMSb}{U}{msb}{m}{n}
\DeclareMathSymbol{\N}{\mathbin}{AMSb}{"4E}
\DeclareMathSymbol{\Z}{\mathbin}{AMSb}{"5A}
\DeclareMathSymbol{\R}{\mathbin}{AMSb}{"52}
\DeclareMathSymbol{\Q}{\mathbin}{AMSb}{"51}
\DeclareMathSymbol{\I}{\mathbin}{AMSb}{"49}
\DeclareMathSymbol{\C}{\mathbin}{AMSb}{"43}

\begin{document}

\title{Optimal fingerprinting strategies with one-sided error}

\author{A. J. Scott}
\email{ascott@qis.ucalgary.ca}
\affiliation{Institute for Quantum Information Science, University of Calgary, Calgary, Alberta T2N 1N4, Canada}

\author{Jonathan Walgate}
\affiliation{Institute for Quantum Information Science, University of Calgary, Calgary, Alberta T2N 1N4, Canada}

\author{Barry C. Sanders}
\affiliation{Institute for Quantum Information Science, University of Calgary, Calgary, Alberta T2N 1N4, Canada}

\begin{abstract}
Fingerprinting enables two parties to infer whether the messages they hold are
the same or different when the cost of communication is high: each message
is associated with a smaller fingerprint and comparisons between messages are
made in terms of their fingerprints alone. In the simultaneous message passing 
model, it is known that fingerprints composed of quantum information can be 
made exponentially smaller than those composed of classical information. For 
small message lengths, we present constructions of optimal classical 
fingerprinting strategies with one-sided error, in both the one-way and 
simultaneous message passing models, and provide bounds on the worst-case 
error probability with the help of extremal set theory. The performance of 
these protocols is then compared to that for quantum fingerprinting strategies 
constructed from spherical codes, equiangular tight frames and mutually unbiased bases.
\end{abstract}

\maketitle

\theoremstyle{plain}
\newtheorem{thm}{Theorem}
\newtheorem*{thmn}{Theorem}
\newtheorem{lem}[thm]{Lemma}
\newtheorem*{lem3p}{Lemma $\bf{3'}$}
\newtheorem{cor}[thm]{Corollary}
\newtheorem{prp}[thm]{Proposition}
\newtheorem{con}[thm]{Conjecture}

\theoremstyle{definition}
\newtheorem{dfn}[thm]{Definition}

\theoremstyle{remark}
\newtheorem*{rmk}{Remark}
\newtheorem*{pprf}{Proof}
\newtheorem{exm}{Example}

\def\rank{\operatorname{rank}}
\def\tr{\operatorname{tr}}
\def\wt{\operatorname{wt}}
\def\d{\operatorname{d}}
\def\ket#1{|#1\rangle}
\def\bra#1{\langle#1|}
\def\braket#1#2{\langle #1 | #2 \rangle}

\def\Pe{P_\text{\rm e}}
\def\Pwce{P_\text{\rm wce}}
\def\Pave{P_\text{\rm ave}}
\def\Pwcem{\mathscr{P}_\text{\rm wce}}
\def\Pavem{\mathscr{P}_\text{\rm ave}}
\def\Pwcemo{\mathscr{P}_\text{\rm wce}^1}
\def\Pavemo{\mathscr{P}_\text{\rm ave}^1}
\def\Is{I}
\def\Ia{\mathcal{I}}

\def\Q#1{\operatorname{Q}(\C^{#1})}

\section{Introduction}

Consider the following scenario. Alice needs to send an important message to Bob. She does not mind if an 
intermediary party reads this message while in transit, but does however, require that the message remains 
unchanged. Alice considers broadcasting her message through a trusted public communication channel, her 
local radio station, but alas, access time is very expensive and her message is long. Instead she 
proceeds by sending it across untrusted communication channels, over the internet, which are 
inexpensive and plentiful. Alice is now faced, however, with the possibility that a saboteur has 
surreptitiously intercepted and changed part, or all, of her message. To test its integrity, she 
reluctantly reconsiders buying airtime at her local radio station. Alice's transmission costs now 
depend on the level of risk she is prepared to tolerate, for a corrupted message to pass undetected. 
An absolute guarantee of integrity requires Alice to transmit her entire message again, this time through 
the trusted channel. If however, she is willing to tolerate a small probability of detection error, a 
considerably smaller piece of information, called a fingerprint, need only be sent. Alice could, for example, 
broadcast only the first byte of her message. This strategy is said to have one-sided error: when the first 
byte does not match that for Bob's copy, Alice knows with certainty that foul play has occurred. Unfortunately, 
since it is Bob who must perform the comparison, and no private communication channel is available, we are 
forced to assume that Alice and Bob use a strategy which is publicly known. An adversary will now simply 
avoid corrupting the first byte, while playing havoc with the rest. There is, of course, a simple 
workaround: Alice instead broadcasts a random byte of her message, chosen by flipping a private coin, 
together with the location of this byte. Assuming the availability of a trusted, though costly, 
broadcast channel, Alice has thus obtained a level of security through only public communication.
In a probabilistic sense, message sabotage will never go undetected. 

Although motivated in terms of message authentication, the preceding protocol is called {\it fingerprinting}.  
In all generality, fingerprinting enables separated parties to infer whether the messages they hold are
the same or different when the cost of communication is high: each message is associated with a smaller 
fingerprint, and comparisons between messages are made in terms of their fingerprints alone. In the above 
authentication scheme, the messages were compared using one-way communication, from Alice to Bob. It is known 
from the theory of communication complexity~\cite{Kushilevitz97,Yao79}, that in the limit of long messages, it is sufficient, 
and necessary, for Alice to communicate fingerprints of length $\Theta(\log N)$ bits, for messages of length $N$ bits, 
if a small constant error probability is allowed. This error probability can then be made arbitrarily small through 
repeated use of the protocol. In another model of communication complexity, called the simultaneous message passing 
(SMP) model~\cite{Yao79}, the fingerprints are transmitted by both Alice and Bob, but this time to a referee, who decides 
the outcome of the message comparison. In this case it is sufficient, and necessary, for Alice and Bob to communicate 
fingerprints of length $\Theta(\sqrt{N})$ bits if the error is to be kept arbitrarily small~\cite{Ambainis96,Newman96,Babai97}.

We will study both communication models by considering the following generalized scenario. A supplier, who we call Sapna, 
chooses two messages, $x$ and $y$, from a pool of $n$ unique messages and hands them to Alice and Bob, respectively. 
Alice and Bob are now tasked with determining whether the messages they hold are the same or different. They are 
forbidden direct communication, however, but instead allowed to correspond with a referee, called Roger. As communication 
is considered expensive, Alice and Bob are limited to sending fingerprints of their original messages, $a$ and $b$ 
respectively, which they select from smaller pools of size $m_A$ and $m_B$. Roger then infers
\begin{equation}
\text{EQ}(x,y)=\left\{\begin{array}{ll} 1,&\text{if }x=y \\ 0,&\text{if }x\neq y\end{array}\right.
\end{equation}
and completes the protocol by announcing a single bit $z\in\{0,1\}$. Roger is correct if 
$z=\text{EQ}(x,y)$. In the current investigation we are concerned only with one-sided-error protocols, in which case, 
$z=0$ is allowed only if $x\neq y$. Then, when Roger announces $0$, Alice and Bob can conclusively claim 
that the messages are indeed different. Protocols with one-sided error are of vital practical importance whenever 
the cost of false negative results outweighs that of false positives. This is the case when $x$ and $y$ are in fact expected to be 
equal. In the above authentication scheme, the one-sided-error condition means that Bob will always accept an uncorrupted 
message as authentic, and thus, transmission throughput remains unaffected when an adversary is not present.

The fingerprinting protocol adopted by Alice, Bob and Roger is publicly announced. The goal
of this protocol is to minimize Roger's error probability. Sapna, however, may be a saboteur,
and always choose message pairs that lead to the highest error rate in Roger's output.
We thus evaluate fingerprinting protocols according to this worst-case scenario. The
worst-case error probability, $\Pwce=\max_{x,y}\text{Prob}(z\neq\text{EQ}(x,y))$, then
corresponds to the maximum error rate, and provides an absolute guarantee on the performance
of the protocol.

Note that the one-way communication model is realized when $m_B=n$, in which case Bob has the same number of 
fingerprints as Sapna has messages, and may simply pass on $y$ to Roger. Bob and Roger may now be thought of 
as the same party. The simultaneous message passing model is realized 
when $m_A=m_B$. We are concerned with the small message limit, however, where it makes more sense to analyze both models 
within the above general scenario. One could also consider a model where two-way communication between 
Alice and Bob is allowed. The added possibility of multiple communication rounds, however, would make the analysis 
of this model considerably more complex. We thus exclude this case.

Quantum fingerprinting protocols \cite{Buh01,Yao03,Niel04,Ambainis03} allow each classical fingerprint to be replaced by a 
quantum state. For a fair comparison, the quantum fingerprints are drawn from a Hilbert space of dimension equal to the 
number of available classical fingerprints. That is, instead of drawing from sets of size $m_A$ and $m_B$, respectively, 
Alice and Bob draw their quantum fingerprints from Hilbert spaces of dimension $m_A$ and $m_B$. In the simultaneous message 
passing model it is known that fingerprints composed of quantum information can be made exponentially smaller than
those composed of classical information. Specifically, for messages of length $N$ bits, it was shown by Buhrman 
{\it et al.}~\cite{Buh01} that fingerprints composed of $\Theta(\log N)$ qubits are sufficient, and necessary, to keep the error 
arbitrarily small. This exponential resource advantage is not apparent in the one-way communication model, where the  
classical and quantum complexities are equal.

In the current article we will investigate optimal fingerprinting strategies in the small message limit.
We will present specific constructions of classical fingerprinting strategies which derive from constant-weight 
codes, and provide lower bounds on the worst-case error probability with the help of extremal set theory. These 
bounds define error rates which quantum protocols must surpass in order to claim a definitive advantage over classical
protocols, and are thus important for current experimental tests of quantum fingerprinting~\cite{Horn04,Du04}. In most 
cases the aforementioned classical fingerprinting strategies are optimal, having error rates that meet a lower bound. 
Our results for classical fingerprinting will be contained in Sec.~\ref{classicalsec}, which is further subdivided into a 
preliminary discussion to set notation, an indepth study of fingerprinting under the one-way communication model, and a study of 
general fingerprinting which concentrates on the SMP model. In Sec.~\ref{quantumsec} we will investigate the extent at which classical strategies 
are outperformed by quantum fingerprinting strategies. Constructions of quantum strategies will be presented which derive 
from spherical codes, equiangular tight frames and mutually unbiased bases. Finally, in Sec.~\ref{concludesec} we summarize our results.

\section{Classical fingerprinting}
\label{classicalsec}
\subsection{Notation and preliminary results}

It will prove useful to think of Alice, Bob and Roger forming a team with the
common goal of minimizing Roger's error probability, and Sapna operating as their
opponent. This team uses a pre-established, publicly known protocol.
In this protocol, Alice and Bob have a fixed probability of communicating each
fingerprint pair $(a,b)$ to Roger for given message pair $(x,y)$ provided by Sapna.
Furthermore, Roger has a fixed probability of declaring $x$ and $y$ to be the 
same message upon receipt of fingerprint pair $(a,b)$.

Any classical fingerprinting protocol is completely specified by three functions:
$p:\{1, \ldots ,m_A\}\times\{1,\ldots ,n\} \rightarrow [0,1]$, 
$q:\{1, \ldots ,m_B\}\times\{1, \ldots ,n\} \rightarrow [0,1]$ and
$r:\{1, \ldots ,m_A\}\times\{1, \ldots ,m_B\}\rightarrow [0,1]$. The function $p(a|x)$ is
the probability that Alice sends fingerprint $a$ to Roger, given that she receives
message $x$ from Sapna. Similarly, $q(b|y)$ is the
probability that Bob sends $b$ to Roger, given that he receives $y$
from Sapna. The function $r(a,b)$ is the probability that Roger outputs $z=1$, 
given that he receives fingerprint $a$ from Alice and $b$ from Bob.

We will call the triple $(p,q,r)$ a {\it strategy}.
When a party's private strategy (i.e. $p$, $q$ or $r$) takes values
only in the set $\{0,1\}$, we call that party's strategy {\it
deterministic}. If all parties' strategies are deterministic we
call the triple $(p,q,r)$ a \emph{deterministic strategy}.
Otherwise a general (i.e. probabilistic) strategy should be
assumed. Normalization requires
\begin{equation}
\sum_{a=1}^{m_A} p(a|x)=\sum_{b=1}^{m_B} q(b|y)=1
\end{equation}
for all $x$ and $y$.

Given a strategy $(p,q,r)$, the
probability that Roger outputs 1 when Sapna deals $x$ to Alice and
$y$ to Bob is
\begin{equation}
P_1^{(p,q,r)}(x,y)\equiv\sum_{a,b} p(a|x)q(b|y)r(a,b)\;.
\label{p1eq}\end{equation}
A strategy is said to have \emph{one-sided error} when
\begin{equation}
P_1^{(p,q,r)}(x,x)=1
\label{onesideconstraint}\end{equation}
for all $x$. Using such a strategy, it is impossible for Roger to announce 0
when Sapna has supplied Alice and Bob with identical messages.

Defining the \emph{error probability}
\begin{equation}
\Pe^{(p,q,r)}(x,y)\equiv\left\{\begin{array}{ll}
1-P_1^{(p,q,r)}(x,x), & x=y \\
P_1^{(p,q,r)}(x,y), & x\neq y\end{array}\right.\;,
\end{equation}
the \emph{average error probability} is then
\begin{equation}
\Pave^{(p,q,r)}\equiv\frac{1}{n^2}\sum_{x,y}\;\Pe^{(p,q,r)}(x,y)\;,
\end{equation}
and the \emph{minimum achievable average error probability} is
\begin{equation}
\Pavem(n,m_A,m_B)\equiv\min_{p,q,r}\;\Pave^{(p,q,r)}\;,
\end{equation}
where the minimum is taken over \emph{all} strategies. If, however, the minimum is restricted to only 
those strategies with one-sided error, we use $\Pavemo(n,m_A,m_B)$ to denote the corresponding minimum achievable 
average error probability. It is this second quantity which will be of interest in the current article. 
The results of Horn {\it et al.} \cite{horn} immediately give us our first result:

\begin{thm}
\begin{equation}
\Pavemo(n,m_A,m_B)=\frac{k \lceil n/m \rceil^2 + (m-k)\lfloor n/m \rfloor^2 - n}{n^2}
\end{equation}
where $\;m=\min\{m_A,m_B\}\;$ and $\;k= n\mod m$.
\label{PaveThm}\end{thm}

When $m_A=m_B=m$ the proof of Theorem~\ref{PaveThm} follows immediately from Lemmas 2 and 3 of \cite{horn}. 
The last paragraph in Sec. 2 of \cite{horn} explains how to deal with the general case. The strategy 
which achieves the optimal error probability, given by Eq. (20) of \cite{horn}, is in fact deterministic. 

Although the average error probability is a perfectly legitimate measure of the performance of a 
fingerprinting strategy, we will consider the worst-case error probability in this article, which corresponds 
to the guaranteed performance of a strategy. The \emph{worst-case error probability} is simply the largest 
error probability that Sapna can coerce:
\begin{equation}
\Pwce^{(p,q,r)}\equiv\max_{x,y}\;\Pe^{(p,q,r)}(x,y)\;.
\end{equation}
Similar to the above, we define the \emph{minimum achievable worst-case error probability}
\begin{equation}
\Pwcem(n,m_A,m_B)\equiv\min_{p,q,r}\;\Pwce^{(p,q,r)}=\min_{p,q,r}\;\max_{x,y}\;\Pe^{(p,q,r)}(x,y)\;,
\end{equation}
and denote by $\Pwcemo(n,m_A,m_B)$ the minimum achievable worst-case error probability when 
restricted to one-sided-error strategies. Our objective in the following sections is to find bounds 
on this second quantity.

To begin, let us introduce a lemma that allows the following simplification.
Whereas Roger can use a probabilistic strategy, we show that there exists a
deterministic strategy for Roger, that is at least as good as all probabilistic 
strategies. 

\begin{lem}
Let $(p,q,r)$ be a fingerprinting strategy with one-sided error. Then
\begin{equation}
\Pe^{(p,q,r)}(x,y)\;\geq\;\Pe^{(p,q,r')}(x,y)
\end{equation}
for all $x$ and $y$, where
\begin{equation}
    r'(a,b)=\left\{\begin{array}{ll}
    1,&\text{if $p(a|x)>0$ and $q(b|x)>0$ for some $x$}\\
    0,&\text{otherwise}\end{array}\right.\;.
\label{roger}\end{equation}
\label{binlem}\end{lem}

The proof of this lemma is straightforward and follows from enforcing the one-sided-error constraint 
[Eq.~(\ref{onesideconstraint})], by setting $r'(a,b)=1$ for each appropriate fingerprint pair $(a,b)$, 
and then choosing $r'(a,b)=0$ in the remaining cases for optimality \cite{horn}. We define a {\it binary strategy} 
$(p,q,r)$ to be one where $r\equiv r'$, as determined by Eq.~(\ref{roger}). All binary strategies have 
one-sided error, and due to Lemma~\ref{binlem}, our search for optimal one-sided-error strategies will be 
limited to this type. Each particular binary matrix $r(a,b)$ can be thought of as defining a class of 
possible strategies for Alice and Bob, i.e. those strategies $p$ and $q$ which do not contradict Eq.~(\ref{roger}). In 
the following sections we will decide which choices for $r(a,b)$ are best. 

\subsection{One-way model: $m_B=n$}
\label{specsec}

The simplest non-trivial examples of classical fingerprinting occur when $m_B=n$, in which case 
Bob may simply pass on $y$ to Roger. In general, when $m_B=n$ we will always set $m_A=m$, 
$q(b|y)=\delta_{by}$, and then refer to the strategy only by the pair $(p,r)$. 
A binary strategy is then one with 
\begin{equation}
r(a,x)=\left\{\begin{array}{ll}
1,&\text{if $p(a|x)>0$}  \\ 
0,&\text{otherwise}\end{array}\right.\;,
\label{binary}\end{equation}
and thus, Alice's decisions will determine Roger's. Alternatively, we may think 
of Roger's binary decision matrix, $r(a,x)$, as specifying a class of strategies with 
$p(a|x)>0$ whenever $r(a,x)=1$, and 0 otherwise. Our job is to then minimize the worst-case error 
probability 
\begin{equation}
\Pwce^{(p,r)}=\max_{x\neq y}\;P_1^{(p,r)}(x,y)
\end{equation}
over all such members of this class, where
\begin{equation}
P_1^{(p,r)}(x,y)=\sum_{a=1}^m p(a|x)r(a,y)\;.
\end{equation}

We first investigate cases where the values of $m$ and $n$ restrict all strategies to those 
with $\Pwce^{(p,r)}=1$. First note that given $\sum_{a}p(a|x)=1$ and $r(a,y)\leq 1$, we 
will have $P_1(x,y)=1$ if and only if $r(a,y)=1$ whenever $p(a|x)>0$, or, for 
binary strategies, $r(a,y)=1$ whenever $r(a,x)=1$. Defining index sets for Alice's 
fingerprinting strategy  
\begin{equation}
\Is_x^{(p)} \equiv \left\{a\,|\,p(a|x)>0\right\}\;,
\end{equation}
which specify Roger's decisions
\begin{equation}
r(a,x)=\left\{\begin{array}{ll}
1,&\text{if $a\in\Is_x^{(p)}$}   \\ 
0,&\text{otherwise}\end{array}\right.\;,
\end{equation}
we have a straightforward lemma.
\begin{lem}
Let $(p,r)$ be a binary fingerprinting strategy. Then $\;\Pwce^{(p,r)}=1\;$ iff $\;\Is_x^{(p)}\subseteq \Is_y^{(p)}$ for some $x\neq y$.
\label{preantilem}\end{lem}
We now restate Lemma~\ref{preantilem} in the language of extremal set theory \cite{anderson,engel,babai}. Roger's decision matrix may be specified 
by the set $\Ia^{(p)}\equiv\big\{\Is_x^{(p)}\,|\,x=1,\dots,n\big\}$ 
which is a subset of the power set $2^{[m]}$, where $[m]\equiv\{1,\dots,m\}$ is called the $m$-set. A $k$-subset of 
a set $X$ is simply a subset of $X$ with cardinality $k$, and the set of all such subsets will be denoted by the symbol ${X\choose k}$. 
Define an {\it antichain} $\mathcal{A}$ of the poset $(\mathcal{S},\subseteq)$ to be a subset of $\mathcal{S}$ with the 
property that no two members are comparable via the relation $\subseteq\,$, i.e. $X\nsubseteq Y$ for all 
distinct $X,Y\in\mathcal{A}$. The {\it length} of the antichain is its cardinality, $|\mathcal{A}|$. 
The following is now equivalent to Lemma~\ref{preantilem}.
\begin{lem3p}
Let $(p,r)$ be a binary fingerprinting strategy. Then $\;\Pwce^{(p,r)}<1\;$ iff $\;\Ia^{(p)}$ is an antichain of length $n$ in $(2^{[m]},\subseteq)$.
\label{antilem}\end{lem3p}
An example of an antichain is $\big\{\{1,2\},\{1,3\},\{1,4\},\{2,3\},\{2,4\},\{3,4\}\big\}$, which is also 
the largest antichain in $2^{[4]}$, and most easily represented through its binary incidence matrix:
\begin{table}
{\footnotesize
\begin{tabular}{|c|@{\hspace{0.3cm}}c@{\hspace{0.3cm}}c@{\hspace{0.3cm}}c@{\hspace{0.3cm}}c@{\hspace{0.3cm}}c@{\hspace{0.3cm}}c@{\hspace{0.3cm}}c@{\hspace{0.3cm}}c@{\hspace{0.3cm}}c@{\hspace{0.3cm}}c@{\hspace{0.3cm}}c@{\hspace{0.3cm}}c@{\hspace{0.3cm}}c@{\hspace{0.3cm}}c@{\hspace{0.3cm}}c@{\hspace{0.3cm}}|}\hline
 $q \backslash m$  & 2 & 3 & 4 & 5 & 6 & 7 & 8 & 9 &10 &11 &12 &13 &14 &15 &16 \\ \hline 
 1 & 2 & 3 & 6 & 10 & 20 & 35 & 70 & 126 & 252 & 462 & 924 & 1716 & 3432 & 6435 & 12870 \\ 
 4/3 & 2 & 3 & 6 & 10 & 20 & 35 & $\geq 56$ & $\geq 84$ & $\geq 120$ & $\geq 165$ & $\geq 220$ & $\geq 286$ & $\geq 364$ & $\geq 455$ & $\geq 560$ \\
 3/2 & 2 & 3 & 6 & 10 & 15 & 21 & $\geq 28$ & $\geq 36$ & $\geq 45$ & $\geq 55$ & $\geq 66$ & $\geq 78$ & $\geq 91$ & $\geq 105$ & $\geq 120$ \\
 2 & 2 & 3 & 4 & 5 & 6 & 7 & 8 & 12 & $13$ & $\geq 17$ & $\geq 20$ & $\geq 26$ & $\geq 28$ & $\geq 35$ & $\geq 37$ \\
 3 & 2 & 3 & 4 & 5 & 6 & 7 & 8 & 9 & 10 & 11 & 12 & 13 & 14 & 15 & $\geq 20$ \\
 4 & 2 & 3 & 4 & 5 & 6 & 7 & 8 & 9 & 10 & 11 & 12 & 13 & 14 & 15 & $\geq 16$ \\\hline
\end{tabular}}
\caption{Known values of $T(m,q)$, the largest $q$-cover free family in $2^{[m]}$, or a lower bound when unknown.} 
\label{qcoverfreetable}\end{table}
\begin{equation}
\begin{matrix}
1 & 2 & 3 & 4 \\\hline
1 & 1 & 0 & 0 \\
1 & 0 & 1 & 0 \\
1 & 0 & 0 & 1 \\
0 & 1 & 1 & 0 \\
0 & 1 & 0 & 1 \\
0 & 0 & 1 & 1 
\end{matrix}\qquad.
\label{antichain46}\end{equation}
In our case this incidence matrix is in fact $r(a,x)$, where $a$ indexes the columns and $x$ the rows. The above 
antichain has length 6. In general, the length of the largest antichain in $2^{[m]}$ is ${m\choose \lfloor m/2 \rfloor}$. 
This result is due to Sperner \cite{sperner} and gives us our first important theorem:
\begin{thm}
$\Pwcemo(n,m,n)<1\;$ iff $\;n\leq {m\choose\lfloor m/2\rfloor}$ .
\label{antibound}\end{thm}
When $m<n\leq{m\choose \lfloor m/2 \rfloor}$ an optimal strategy can be found by first collecting all 
antichains of length $n$, and then for each such candidate for $\Ia^{(p)}$, minimize $\Pwce^{(p,r)}$ over 
Alice's transition probabilities $p(a|x)$ to locate the optimal choice. Consider the case when 
$n={m\choose \lfloor m/2 \rfloor}$ for example. If $m$ is even, the set of 
all $m/2$-subsets of the $m$-set, ${[m]\choose m/2}$, is the only antichain; if 
$m$ is odd, the sets ${[m]\choose (m-1)/2}$ and ${[m]\choose (m+1)/2}$ are the only antichains \cite{anderson,lovasz}. 
In both cases an optimal strategy follows from choosing $\Ia^{(p)}=\big\{\Is_x^{(p)}\,|\,x=1,\dots,n\big\}={[m]\choose \lfloor m/2 \rfloor}$ 
and then $p(a|x)=1/\lfloor m/2 \rfloor$ when $a\in\Is_x^{(p)}$ (and zero otherwise). For example, when 
$m=4$ and $n={4\choose 2}=6$ the above antichain [Eq.~(\ref{antichain46})] is the only 
choice, in which case setting $p(a|x)=r(a,x)/2$ is optimal, resulting in $\Pwce^{(p,r)}=1/2$. In general, the 
resulting worst-case error probability is stated in the following proposition. 
\begin{prp}
$\Pwcemo(n,m,n)=1-1/\lfloor m/2\rfloor\;$ if $\;n={m\choose\lfloor m/2\rfloor}$ .
\end{prp}
Searching for antichains is computationally expensive except for small $m$ and $n$. We instead turn our 
attention to deriving upper and lower bounds for $\Pwcemo(n,m,n)$. Lower bounds follow from generalizing the 
concept of an antichain to a $k$-cover free family of sets \cite{erdos,erdos2}. A {\it family} of sets is 
simply a set of sets. We call the family $\mathcal{F}$, {\it $k$-cover free} when 
$X\nsubseteq Y_1\cup\dots\cup Y_k$ for all $X,Y_1,\dots,Y_k\in\mathcal{F}$, where $X\neq Y_i$ for all $i$. 
Note that a $1$-cover free family is an antichain. The following incidence matrix specifies an example of 
a $2$-cover free family. 
\begin{equation}
\begin{matrix}
1 & 2 & 3 & 4 & 5 & 6 & 7 & 8 & 9 \\\hline
1 & 1 & 1 & 0 & 0 & 0 & 0 & 0 & 0 \\
0 & 0 & 0 & 1 & 1 & 1 & 0 & 0 & 0 \\
0 & 0 & 0 & 0 & 0 & 0 & 1 & 1 & 1 \\
1 & 0 & 0 & 1 & 0 & 0 & 1 & 0 & 0 \\
0 & 1 & 0 & 0 & 1 & 0 & 0 & 1 & 0 \\
0 & 0 & 1 & 0 & 0 & 1 & 0 & 0 & 1 \\
1 & 0 & 0 & 0 & 1 & 0 & 0 & 0 & 1 \\
0 & 1 & 0 & 0 & 0 & 1 & 1 & 0 & 0 \\
0 & 0 & 1 & 1 & 0 & 0 & 0 & 1 & 0 \\
1 & 0 & 0 & 0 & 0 & 1 & 0 & 1 & 0 \\
0 & 1 & 0 & 1 & 0 & 0 & 0 & 0 & 1 \\
0 & 0 & 1 & 0 & 1 & 0 & 1 & 0 & 0 
\end{matrix}
\label{coverfree912}\end{equation}

Suppose that Alice uses a strategy where $\Is_x^{(p)}\subseteq\Is_y^{(p)}\cup\Is_z^{(p)}$
for some $x\neq y,z$. Then, for one choice between the message pairs $(x,y)$ and $(x,z)$, 
Sapna will be able to coerce an error probability of 1/2 or greater. This follows from the inequality
\begin{equation}
P_1^{(p,r)}(x,y)+P_1^{(p,r)}(x,z) \;=\; \sum_{a=1}^m p(a|x)\left[r(a,y)+r(a,z)\right] \;\geq\; \sum_{a=1}^m p(a|x)r(a,x)\;=\;1 \;,
\label{simpleboundequation}\end{equation} 
since the assumption $\Is_x^{(p)}\subseteq\Is_y^{(p)}\cup\Is_z^{(p)}$ 
implies that either $r(a,y)=1$ or $r(a,z)=1$ when $r(a,x)=1$. Thus the worst-case error probability 
may be less than 1/2 only when $\Ia^{(p)}$ is a $2$-cover free family. This observation, in its generalized 
form, is the content of the following lemma. 

\begin{lem}
Let $(p,r)$ be a binary fingerprinting strategy. Then $\;\Pwce^{(p,r)}<1/k\;$ only if $\;\Ia^{(p)}$ is a $k$-cover free family of size $n$ in $2^{[m]}$.
\label{lemmaCoverFree}\end{lem}

Unlike in the case of antichains, the size of the largest $k$-cover free family in the power set $2^{[m]}$ 
is a difficult problem, and consequently, unknown except in a few special cases. Nonetheless, let us 
denote this number by $T(m,k)$. The following theorem then applies.

\begin{thm}
$\Pwcemo(n,m,n)\geq 1/k\;$ if $\;n>T(m,k)\;$.
\label{coverfreebound}\end{thm}

\begin{cor}
$\Pwcemo(n,m,n)\geq 1/k\;$ if $\;{n\choose k}>{m\choose\lfloor m/2\rfloor}\;$.
\end{cor}

The corollary follows from the fact that a family $\mathcal{F}$ is $k$-cover free only if 
the set $\big\{Y_1\cup\dots\cup Y_k\,|\text{ distinct } Y_1,\dots,Y_k \in\mathcal{F}\big\}$ forms an antichain.
The obtained bound, however, may be tightened by calculating the exact value of $T(m,k)$. This problem is also 
known under the name of {\it superimposed codes} \cite{kautz}. In fact, each $k$-cover free family of sets 
$\mathcal{F}$ is equivalent to a superimposed code, in which the binary incidence vectors corresponding to 
members of $\mathcal{F}$ form the codewords. The exact value of $T(m,k)$ is known when $k=1$, in which 
case Sperner's result for the largest antichain applies, and for the following special cases. 
Tables from \cite{kim} show that $T(m,2)=m$ when $m\leq 8$, and $T(m,k)=m$ when $m\leq 15$ and $k\geq 3$. 
Also, our own computational searches have found that $T(9,2)=12$ and $T(10,2)=13$. These results are 
tabulated in Table~\ref{qcoverfreetable} (integral $q$), and give many of the lower bounds in 
Table~\ref{Pwcemotable} for $\Pwcemo(n,m,n)$.

We can generalize Theorem~\ref{coverfreebound} by introducing the concept of a multiset. A {\it multiset} 
is like a set, except that repeated elements are now allowed, e.g. $X=\{1,1,2,3\}$. Multiset operations 
then differ from set operations in a straightforward way: $X\cup\{2\}=\{1,1,2,2,3\}$, $X\cap\{1,1,2,2\}=\{1,1,2\}$, 
and $\{1,1\}\subseteq X$ but $\{1,1,1\}\nsubseteq X$. Let $\cup^j X$ denote the multiset union of $j$ copies of $X$. A family of 
sets $\mathcal{F}$ will be called {\it $k/j$-cover free} when $\cup^j X \nsubseteq Y_1\cup\dots\cup Y_k$, under multiset operations, 
for all $X,Y_1,\dots,Y_k\in\mathcal{F}$, where $X\neq Y_i$ for all $i$. Note that a family of sets is $q$-cover free 
for all rationals $q\leq p$ whenever it is $p$-cover free. An example of a $3/2$-cover free family is 
the set ${[6]\choose 2}$. 

\begin{lem}
Let $(p,r)$ be a binary fingerprinting strategy. Then $\;\Pwce^{(p,r)}<1/q\;$ only if $\;\Ia^{(p)}$ is a $q$-cover free family of size 
$n$ in $2^{[m]}$.
\label{lemmaCoverFree2}\end{lem}

\begin{table}
{\footnotesize
\begin{tabular}{|c|@{\hspace{0.3cm}}c@{\hspace{0.3cm}}c@{\hspace{0.3cm}}c@{\hspace{0.3cm}}c@{\hspace{0.3cm}}c@{\hspace{0.3cm}}c@{\hspace{0.3cm}}c@{\hspace{0.3cm}}c@{\hspace{0.3cm}}c@{\hspace{0.3cm}}c@{\hspace{0.3cm}}c@{\hspace{0.3cm}}c@{\hspace{0.3cm}}c@{\hspace{0.3cm}}c@{\hspace{0.3cm}}c@{\hspace{0.3cm}}|}\hline
  $n\backslash m$ & 2 & 3 & 4 & 5 & 6 & 7 & 8 & 9 &10 &11 &12 &13 &14 &15 &16 \\ \hline 
 2 & 0 & 0 & 0 & 0 & 0 & 0 & 0 & 0 & 0 & 0 & 0 & 0 & 0 & 0 & 0 \\
 3 & 1 & 0 & 0 & 0 & 0 & 0 & 0 & 0 & 0 & 0 & 0 & 0 & 0 & 0 & 0 \\
 4 & 1 & 1 & 0 & 0 & 0 & 0 & 0 & 0 & 0 & 0 & 0 & 0 & 0 & 0 & 0 \\ 
 5 & 1 & 1 & 1/2 & 0 & 0 & 0 & 0 & 0 & 0 & 0 & 0 & 0 & 0 & 0 & 0 \\
 6 & 1 & 1 & 1/2 & 1/2 & 0 & 0 & 0 & 0 & 0 & 0 & 0 & 0 & 0 & 0 & 0 \\
 7 & 1 & 1 & 1 & 1/2 & 1/2 & 0 & 0 & 0 & 0 & 0 & 0 & 0 & 0 & 0 & 0 \\
 8 & 1 & 1 & 1 & 1/2 & 1/2 & 1/2 & 0 & 0 & 0 & 0 & 0 & 0 & 0 & 0 & 0 \\
 9 & 1 & 1 & 1 & 1/2 & 1/2 & 1/2 & 1/2 & 0 & 0 & 0 & 0 & 0 & 0 & 0 & 0 \\
10 & 1 & 1 & 1 & 1/2 & 1/2 & 1/2 & 1/2 & 1/3 & 0 & 0 & 0 & 0 & 0 & 0 & 0 \\
11 & 1 & 1 & 1 & 1 & 1/2 & 1/2 & 1/2 & 1/3 & 1/3 & 0 & 0 & 0 & 0 & 0 & 0 \\
12 & 1 & 1 & 1 & 1 & 1/2 & 1/2 & 1/2 & 1/3 & 1/3 & 1/3 & 0 & 0 & 0 & 0 & 0 \\
13 & 1 & 1 & 1 & 1 & 1/2 & 1/2 & 1/2 & 1/2 & 1/3 & 1/3 & 1/3 & 0 & 0 & 0 & 0 \\
14 & 1 & 1 & 1 & 1 & 1/2 & 1/2 & 1/2 & 1/2 & 1/2 & 1/3 & 1/3 & 1/3 & 0 & 0 & 0 \\
15 & 1 & 1 & 1 & 1 & 1/2 & 1/2 & 1/2 & 1/2 & 1/2 & 1/3 & 1/3 & 1/3 & 1/3 & 0 & 0 \\
16 & 1 & 1 & 1 & 1 & 2/3 & 1/2 & 1/2 & 1/2 & 1/2 & 1/3 & 1/3 & 1/3 & 1/3 & 1/3 & 0 \\
17 & 1 & 1 & 1 & 1 & 2/3 & 1/2 & 1/2 & 1/2 & 1/2 & 1/3 & 1/3 & 1/3 & 1/3 & 1/3 & 0 -- 1/4 \\
18 & 1 & 1 & 1 & 1 & 2/3 & 1/2 & 1/2 & 1/2 & 1/2 & 1/3 -- 1/2 & 1/3 & 1/3 & 1/3 & 1/3 & 0 -- 1/4 \\
19 & 1 & 1 & 1 & 1 & 2/3 & 1/2 & 1/2 & 1/2 & 1/2 & 1/3 -- 1/2 & 1/3 & 1/3 & 1/3 & 1/3 & 0 -- 1/4 \\
20 & 1 & 1 & 1 & 1 & 2/3 & 1/2 & 1/2 & 1/2 & 1/2 & 1/3 -- 1/2 & 1/3 & 1/3 & 1/3 & 1/3 & 0 -- 1/4 \\
21 & 1 & 1 & 1 & 1 & 1 & 1/2 & 1/2 & 1/2 & 1/2 & 1/3 -- 1/2 & 1/3 -- 1/2 & 1/3 & 1/3 & 1/3 & 0 -- 1/3 \\
22 & 1 & 1 & 1 & 1 & 1 & 2/3 & 1/2 & 1/2 & 1/2 & 1/3 -- 1/2 & 1/3 -- 1/2 & 1/3 & 1/3 & 1/3 & 0 -- 1/3 \\
23 & 1 & 1 & 1 & 1 & 1 & 2/3 & 1/2 & 1/2 & 1/2 & 1/3 -- 1/2 & 1/3 -- 1/2 & 1/3 & 1/3 & 1/3 & 0 -- 1/3 \\
24 & 1 & 1 & 1 & 1 & 1 & 2/3 & 1/2 & 1/2 & 1/2 & 1/3 -- 1/2 & 1/3 -- 1/2 & 1/3 & 1/3 & 1/3 & 0 -- 1/3 \\
25 & 1 & 1 & 1 & 1 & 1 & 2/3 & 1/2 & 1/2 & 1/2 & 1/3 -- 1/2 & 1/3 -- 1/2 & 1/3 & 1/3 & 1/3 & 0 -- 1/3 \\
26 & 1 & 1 & 1 & 1 & 1 & 2/3 & 1/2 & 1/2 & 1/2 & 1/3 -- 1/2 & 1/3 -- 1/2 & 1/3 & 1/3 & 1/3 & 0 -- 1/3 \\
27 & 1 & 1 & 1 & 1 & 1 & 2/3 & 1/2 & 1/2 & 1/2 & 1/3 -- 1/2 & 1/3 -- 1/2 & 1/3 -- 1/2 & 1/3 & 1/3 & 0 -- 1/3 \\
28 & 1 & 1 & 1 & 1 & 1 & 2/3 & 1/2 & 1/2 & 1/2 & 1/3 -- 1/2 & 1/3 -- 1/2 & 1/3 -- 1/2 & 1/3 & 1/3 & 0 -- 1/3 \\
29 & 1 & 1 & 1 & 1 & 1 & 2/3 & 1/2 -- 2/3 & 1/2 & 1/2 & 1/3 -- 1/2 & 1/3 -- 1/2 & 1/3 -- 1/2 & 1/3 -- 1/2 & 1/3 & 0 -- 1/3 \\
30 & 1 & 1 & 1 & 1 & 1 & 2/3 & 1/2 -- 2/3 & 1/2 & 1/2 & 1/3 -- 1/2 & 1/3 -- 1/2 & 1/3 -- 1/2 & 1/3 -- 1/2 & 1/3 & 0 -- 1/3 \\
31 & 1 & 1 & 1 & 1 & 1 & 2/3 & 1/2 -- 2/3 & 1/2 & 1/2 & 1/3 -- 1/2 & 1/3 -- 1/2 & 1/3 -- 1/2 & 1/3 -- 1/2 & 1/3 & 0 -- 1/3 \\
32 & 1 & 1 & 1 & 1 & 1 & 2/3 & 1/2 -- 2/3 & 1/2 & 1/2 & 1/3 -- 1/2 & 1/3 -- 1/2 & 1/3 -- 1/2 & 1/3 -- 1/2 & 1/3 & 0 -- 1/3 \\
33 & 1 & 1 & 1 & 1 & 1 & 2/3 & 1/2 -- 2/3 & 1/2 & 1/2 & 1/3 -- 1/2 & 1/3 -- 1/2 & 1/3 -- 1/2 & 1/3 -- 1/2 & 1/3 & 0 -- 1/3 \\
34 & 1 & 1 & 1 & 1 & 1 & 2/3 & 1/2 -- 2/3 & 1/2 & 1/2 & 1/3 -- 1/2 & 1/3 -- 1/2 & 1/3 -- 1/2 & 1/3 -- 1/2 & 1/3 & 0 -- 1/3 \\
35 & 1 & 1 & 1 & 1 & 1 & 2/3 & 1/2 -- 2/3 & 1/2 & 1/2 & 1/3 -- 1/2 & 1/3 -- 1/2 & 1/3 -- 1/2 & 1/3 -- 1/2 & 1/3 & 0 -- 1/3 \\
36 & 1 & 1 & 1 & 1 & 1 & 1 & 1/2 -- 2/3 & 1/2 & 1/2 & 1/3 -- 1/2 & 1/3 -- 1/2 & 1/3 -- 1/2 & 1/3 -- 1/2 & 1/3 -- 1/2 & 0 -- 1/3 \\
37 & 1 & 1 & 1 & 1 & 1 & 1 & 1/2 -- 2/3 & 1/2 -- 2/3 & 1/2 & 1/3 -- 1/2 & 1/3 -- 1/2 & 1/3 -- 1/2 & 1/3 -- 1/2 & 1/3 -- 1/2 & 0 -- 1/3 \\
38 & 1 & 1 & 1 & 1 & 1 & 1 & 1/2 -- 2/3 & 1/2 -- 2/3 & 1/2 & 1/3 -- 1/2 & 1/3 -- 1/2 & 1/3 -- 1/2 & 1/3 -- 1/2 & 1/3 -- 1/2 & 0 -- 1/2 \\
39 & 1 & 1 & 1 & 1 & 1 & 1 & 1/2 -- 2/3 & 1/2 -- 2/3 & 1/2 & 1/3 -- 1/2 & 1/3 -- 1/2 & 1/3 -- 1/2 & 1/3 -- 1/2 & 1/3 -- 1/2 & 0 -- 1/2 \\
40 & 1 & 1 & 1 & 1 & 1 & 1 & 1/2 -- 2/3 & 1/2 -- 2/3 & 1/2 & 1/3 -- 1/2 & 1/3 -- 1/2 & 1/3 -- 1/2 & 1/3 -- 1/2 & 1/3 -- 1/2 & 0 -- 1/2 \\\hline
\end{tabular}}
\caption{$\Pwcemo(n,m,n)$, the minimum achievable worst-case error probability for a classical fingerprinting strategy under 
the one-way communication model, or the range of possible values when unknown.} 
\label{Pwcemotable}\end{table}

This lemma follows from a straightforward generalization of Eq.~(\ref{simpleboundequation}):
Suppose now that $\cup^j\Is_x^{(p)}\subseteq\Is_{y_1}^{(p)}\cup\dots\cup\Is_{y_k}^{(p)}$, 
under multiset operations, for some $x\neq y_1,\dots,y_k$. Then
\begin{eqnarray}
P_1^{(p,r)}(x,y_1)+\dots+P_1^{(p,r)}(x,y_k) &=& \sum_{a=1}^m p(a|x)\left[r(a,y_1)+\dots+r(a,y_k)\right] \\
 &\geq& j\sum_{a=1}^m p(a|x)r(a,x) \\
 &=& j \;,
\label{notassimpleboundequation}\end{eqnarray} 
since the condition $\cup^j\Is_x^{(p)}\subseteq\Is_{y_1}^{(p)}\cup\dots\cup\Is_{y_k}^{(p)}$
means that $r(a,y_l)=1$ whenever $r(a,x)=1$ for $j$ different choices of $1\leq l \leq k$. 
Equation~(\ref{notassimpleboundequation}) implies that $P_1^{(p,r)}(x,y_l)\geq j/k$ for at least one 
$l$, and thus the worst-case error probability can be less than $j/k$ only when 
$\Ia^{(p)}$ is a $k/j$-cover free family. 

For rational $q$, if we let $T(m,q)$ denote the size of the largest $q$-cover free family in the power 
set $2^{[m]}$, then Theorem~\ref{coverfreebound} generalizes to the following. 

\begin{thm}
$\Pwcemo(n,m,n)\geq 1/q\;$ if $\;n>T(m,q)\;$.
\label{coverfreebound2}\end{thm}

Our own computational searches have found that $T(m,3/2)={m\choose 2}$ for $m\leq 7$, which is also 
tabulated in Table~\ref{qcoverfreetable}, and gives the corresponding lower bounds on $\Pwcemo(n,m,n)$ 
in Table~\ref{Pwcemotable}. This formula, however, is unlikely to be true in general.

We obtain an upper bound on $\Pwcemo(n,m,n)$ whenever a particular example strategy is found. Consider 
those derived from {\it constant weight codes} \cite{brouwer,agrell}. Recall that the weight of a 
binary vector $\bm{u}=\{u_i\}_{i=1}^n$ is $\wt(\bm{u})=\sum_i u_i$, and the Hamming distance 
between two vectors $\bm{u}$ and $\bm{v}$ is $\d(\bm{u},\bm{v})=\sum_i |u_i- v_i|$. The code-theoretic 
function $A(n,d,w)$ is defined as the maximum possible number of binary vectors of length $n$, 
Hamming distance at least $d$ apart, and constant weight $w$. If we let $N(m,k,j)$ denote the maximum 
number of $k$-subsets of the $m$-set with pairwise intersections containing $j$ elements or 
less, then $N(m,k,j)=A(m,2(k-j),k)$. We can see this by considering the codewords as incidence vectors for subsets of 
the $m$-set, thus defining the isomorphism $f:\{0,1\}^m\rightarrow 2^{[m]}$ by $f(\bm{u})=\{i\,|\,u_i=1\}$, and noting that 
$\d(\bm{u},\bm{v})=2(k-|f(\bm{u})\cap f(\bm{v})|)$ when $\wt(\bm{u})=\wt(\bm{v})=k$. Now given a family 
$\mathcal{F}$, of $n$ $k$-subsets of the $m$-set with pairwise intersections containing $j$ elements or less, 
we may construct a strategy with $\Pwce^{(p,r)}\leq j/k$ by uniquely assigning each index matrix $\Is_x^{(p)}$ 
($x=1,..,n$) with a member of $\mathcal{F}$ and setting $p(a|x)=1/k$ when $a\in\Is_x^{(p)}$ (and zero otherwise). 
Thus $\Pwcemo(n,m,n)\leq j/k$ whenever $n\leq N(m,k,j)=A(m,2(k-j),k)$, which is restated in the following 
theorem.

\begin{thm}
$\Pwcemo(n,m,n)\leq j/k\;$ if $\;n\leq N(m,k,j)=A(m,2(k-j),k)\;$.
\label{constantweightbound}\end{thm}

In some cases $A(n,d,w)$ is known exactly \cite{brouwer}. Examples are $N(m,k,k-1)=A(m,2,k)={m \choose k}$, 
which is trivial,
\begin{equation}
N(m,3,1)=A(m,4,3)=\left\{\begin{array}{ll}
\Big\lfloor\frac{m}{3}\Big\lfloor\frac{m-1}{2}\Big\rfloor\Big\rfloor, & \text{if $m\not\equiv 5 \mod 6$} \\ \\
\Big\lfloor\frac{m}{3}\Big\lfloor\frac{m-1}{2}\Big\rfloor\Big\rfloor-1, & \text{if $m\equiv 5 \mod 6$}  
\end{array}\right.
\end{equation}
and
\begin{equation}
N(q^2,q,1)=A(q^2,2q-2,q) = q(q+1)
\label{classicalMUBs}\end{equation}
if $q$ is a prime power. These results give the upper bounds on $\Pwcemo(n,m,n)$ in Table~\ref{Pwcemotable}. Thus far,
we do not know of a strategy with smaller worst-case error probability than that attained by a 
constant-weight-code strategy. 

Finally, we remark that the inequalities 
$\Pwcemo(n+1,m,n+1)\geq\Pwcemo(n,m,n)\geq\Pwcemo(n+1,m+1,n+1)$ trivially hold, since increasing the number 
of messages will not allow lower error probabilities, and a strategy with $n+1$ messages and $m+1$ 
fingerprints can be constructed from one with $n$ messages and $m$ fingerprints, by simply allowing 
Alice to pass on the additional message to Roger as the additional fingerprint.

\subsection{Generalizations and the SMP model: $m_A=m_B$}

We now apply some of the ideas of the previous section to the general case where Bob's fingerprint set 
is also restricted in size. Defining index sets for Alice's and Bob's fingerprinting strategies  
\begin{equation}
\Is_x^{(p)} \equiv \left\{a\,|\,p(a|x)>0\right\}\;,\qquad\Is_y^{(q)} \equiv \left\{b\,|\,q(b|y)>0\right\}\;,
\end{equation} 
a binary strategy is one with 
\begin{equation}
r(a,b)=\left\{\begin{array}{ll}
1,&\text{if $(a,b)\in\cup_z\,\Is_z^{(p)}\times\Is_z^{(q)}$} \\
0,&\text{otherwise}\end{array}\right.\;.
\label{binary2}\end{equation}
After inspection of Eq.~(\ref{p1eq}) it is straightforward to see that Lemma~\ref{preantilem} of the 
previous section takes the following generalized form.
\begin{lem}
Let $(p,q,r)$ be a binary fingerprinting strategy. Then $\Pwce^{(p,q,r)}=1\;$ iff $\;\Is_x^{(p)}\times\Is_y^{(q)}\:\subseteq\: \cup_z\:\Is_z^{(p)}\times\Is_z^{(q)}$ for some $x\neq y$.
\label{lemmaB1}\end{lem}
A necessary condition for $\Pwce^{(p,q,r)}<1$ is to have both families of sets 
$\Ia^{(p)}=\big\{\Is_x^{(p)}\big\}$ and $\Ia^{(q)}=\big\{\Is_y^{(q)}\big\}$
forming antichains. Note however, that a different labelling of members in one of 
the families (through subscript $x$ or $y$) will, in general, correspond to a different strategy; 
consequently, the enumeration of elements in $\Ia^{(p)}$ and $\Ia^{(q)}$ is now important. 

Lemma~\ref{lemmaB1} may be used to derive conditions upon which the worst-case error probability is 
necessarily 1, thus generalizing Theorem~\ref{antibound} of the previous section. We may also generalize 
Theorem~\ref{antibound}, however, in the following straightforward manner. Setting $m=\min\{m_A,m_B\}$, we first note that Theorem~\ref{antibound} 
implies $\Pwcemo(n,m_A,m_B)=1$ whenever $n>{m\choose \lfloor m/2 \rfloor}$, since the minimum possible 
worst-case error probability cannot decrease when the size of Bob's message pool, $m_B$, is decreased. 
We now claim that in fact the converse is also true. That is, whenever $\Pwcemo(n,m_A,m_B)=1$ we must have 
$n>{m\choose \lfloor m/2 \rfloor}$, or equivalently, $\Pwcemo(n,m,m)<1$ when $n={m\choose \lfloor m/2 \rfloor}$. 
Explicit examples of such strategies will prove our claim. Consider the binary strategies 
with  $\Ia^{(p)}={[m]\choose \lfloor m/2 \rfloor}$ (any labelling of elements in this family will do) 
and then $\Is_y^{(q)}\equiv{\Is_y^{(p)}}'$, the complement of $\Is_y^{(p)}$ in $[m]$. Roger's decisions 
are dictated by Eq.~(\ref{binary2}), and hence, $r(a,b)=1-\delta(a,b)$. Finally by choosing, 
for example, $p(a|x)=1/\lfloor m/2 \rfloor$ when $a\in\Is_x^{(p)}$ (otherwise zero), and 
$q(b|y)=1/\lceil m/2\rceil$ when $b\in\Is_y^{(q)}={\Is_y^{(p)}}'$, we find that 
$\Pwce^{(p,q,r)}=1-1/(\lfloor m/2\rfloor\lceil m/2\rceil)<1$ and the following theorem is proven.
\begin{thm}
$\Pwcemo(n,m_A,m_B)<1\;$ iff $\;n\leq {m\choose\lfloor m/2\rfloor}\;$ where $\;m=\min\{m_A,m_B\}\;$  .
\label{antibound2}\end{thm}
The generalization of other results from the previous section was found to be quite difficult. 
We can say some more, however, about the special case of Alice and Bob sharing the same set of 
fingerprints, i.e. when $m_A=m_B=m$. In particular, the strategy given above can be shown to be optimal: 
\begin{prp}
$\Pwcemo(n,m,m)=1-1/(\lfloor m/2\rfloor\lceil m/2\rceil)\;$ if $\;n={m\choose\lfloor m/2\rfloor}$ .
\end{prp}
To see how to show this, consider the case when $m$ is even. 
Since $n={m\choose m/2}$, the only choice for $\Ia^{(p)}$ and $\Ia^{(q)}$ is the antichain ${[m]\choose m/2}$. 
We must now decide how to optimally label the index sets $\Is_x^{(p)}\in\Ia^{(p)}$ and 
$\Is_y^{(q)}\in\Ia^{(q)}$. There is in fact only one choice, modulo permutations of message and fingerprint 
labels, for which $\Is_x^{(p)}\times\Is_y^{(q)}\:\nsubseteq\: \cup_z\:\Is_z^{(p)}\times\Is_z^{(q)}$ for 
every $x\neq y$, and by Lemma \ref{lemmaB1}, allows a worst-case error probability less than 1. One can show that for any labelling of 
the index sets, every row and every column of the $m\times m$ matrix $r(a,b)$ has at least $m-1$ entries 
equal to 1. The condition for the worst-case error probability to be less than 1 can be 
satisfied only for those matrices with a 0 in every row and every column i.e. $r(a,b)$ is a row/column 
permutation of the $m\times m$ matrix with a diagonal of 0's, and off-diagonal elements equal to 1. Such 
strategies correspond to a message and/or fingerprint relabelling of the strategy given above, and are 
optimal when the nonzero transmission probabilities of $p(b|x)$ and $q(b|y)$ are symmetrically chosen to be 
$2/m$. The case when $m$ is odd is similar.

In principle, we can also use Lemma~\ref{lemmaB1} in this way to locate an optimal strategy for other  
$n$. One must, however, test all possible choices for $\Ia^{(p)}$ and $\Ia^{(q)}$, i.e., all 
antichains of length $n$, which grow exponentially in number as $n$ decreases. Consequently, this laborious 
task was undertaken only for two special cases, with results: $\Pwcemo(5,4,4)=3/4$ and $\Pwcemo(9,5,5)=5/6$. 

\begin{table}
{\small
\begin{tabular}{|c|@{\hspace{0.3cm}}c@{\hspace{0.3cm}}c@{\hspace{0.3cm}}c@{\hspace{0.3cm}}c@{\hspace{0.3cm}}c@{\hspace{0.3cm}}c@{\hspace{0.3cm}}c@{\hspace{0.3cm}}c@{\hspace{0.3cm}}|}\hline
  $n\backslash m$ & 2 & 3 & 4 & 5 & 6 & 7 & 8 & 9  \\ \hline 
 2 & 0 & 0 & 0 & 0 & 0 & 0 & 0 & 0  \\
 3 & 1 & 0 & 0 & 0 & 0 & 0 & 0 & 0  \\
 4 & 1 & 1 & 0 & 0 & 0 & 0 & 0 & 0  \\ 
 5 & 1 & 1 & 3/4 & 0 & 0 & 0 & 0 & 0  \\
 6 & 1 & 1 & 3/4 & 1/2 -- 3/4 & 0 & 0 & 0 & 0 \\
 7 & 1 & 1 & 1 & 1/2 -- 3/4 & 1/2 -- 3/4 & 0 & 0 & 0 \\
 8 & 1 & 1 & 1 & 1/2 -- 3/4 & 1/2 -- 3/4 & 1/2 -- 3/4 & 0 & 0 \\
 9 & 1 & 1 & 1 & 5/6 & 1/2 -- 3/4 & 1/2 -- 3/4 & 1/2 -- 3/4 & 0 \\
10 & 1 & 1 & 1 & 5/6 & 1/2 -- 3/4 & 1/2 -- 3/4 & 1/2 -- 3/4 & 1/3 -- 1/2 \\
11 & 1 & 1 & 1 & 1 & 1/2 -- 3/4 & 1/2 -- 3/4 & 1/2 -- 3/4 & 1/3 -- 3/4 \\
12 & 1 & 1 & 1 & 1 & 1/2 -- 3/4 & 1/2 -- 3/4 & 1/2 -- 3/4 & 1/3 -- 3/4 \\
13 & 1 & 1 & 1 & 1 & 1/2 -- 5/6 & 1/2 -- 3/4 & 1/2 -- 3/4 & 1/2 -- 3/4 \\
14 & 1 & 1 & 1 & 1 & 1/2 -- 8/9 & 1/2 -- 3/4 & 1/2 -- 3/4 & 1/2 -- 3/4 \\
15 & 1 & 1 & 1 & 1 & 1/2 -- 8/9 & 1/2 -- 3/4 & 1/2 -- 3/4 & 1/2 -- 3/4  \\
16 & 1 & 1 & 1 & 1 & 2/3 -- 8/9 & 1/2 -- 3/4 & 1/2 -- 3/4 & 1/2 -- 3/4  \\
17 & 1 & 1 & 1 & 1 & 2/3 -- 8/9 & 1/2 -- 3/4 & 1/2 -- 3/4 & 1/2 -- 3/4  \\
18 & 1 & 1 & 1 & 1 & 2/3 -- 8/9 & 1/2 -- 3/4 & 1/2 -- 3/4 & 1/2 -- 3/4  \\
19 & 1 & 1 & 1 & 1 & 2/3 -- 8/9 & 1/2 -- 3/4 & 1/2 -- 3/4 & 1/2 -- 3/4  \\
20 & 1 & 1 & 1 & 1 & 8/9 & 1/2 -- 3/4 & 1/2 -- 3/4 & 1/2 -- 3/4  \\
21 & 1 & 1 & 1 & 1 & 1 & 1/2 -- 3/4 & 1/2 -- 3/4 & 1/2 -- 3/4 \\
22 & 1 & 1 & 1 & 1 & 1 & 2/3 -- 8/9 & 1/2 -- 3/4 & 1/2 -- 3/4 \\\hline
\end{tabular}}
\caption{$\Pwcemo(n,m,m)$, the minimum achievable worst-case error probability for a classical fingerprinting strategy under 
the SMP model, or the range of possible values when unknown.} 
\label{Pwcemotable2}\end{table}

In Table~\ref{Pwcemotable2} we report the known values of $\Pwcemo(n,m,m)$, or the range of possible values 
when unknown. Apart from the above special cases where the exact value of $\Pwcemo(n,m,m)$ was 
derived, all lower bounds are the same as those given in the previous section, being derived from the 
inequality $\Pwcemo(n,m,m)\geq\Pwcemo(n,m,n)$. Such bounds are likely to be weak in general. 

The upper bounds on $\Pwcemo(n,m,m)$ in Table~\ref{Pwcemotable2} come from explicit strategies. First 
note that we will always have $\Pwcemo(2n,m+n,m+n)\leq\Pwcemo(n,m,n)$. This follows from a straightforward 
conversion of an $(n,m,n)$-fingerprinting strategy of the previous section, denoted by
$(p',r')$ say, into a strategy suitable for employment in the current scenario: Previously, Alice 
performed the fingerprinting strategy $p'$, while Bob simply acted as a relay, and passed his 
message on to Roger. To covert this strategy into one suitable for the current scenario, we ask Alice to 
employ $p'$ on the first half of Sapna's messages and relay the second half, while Bob is tasked with the 
opposite, fingerprinting on the second half and relaying on the first. Explicitly,   
\begin{eqnarray}
p(a|x) &=& \left\{\begin{array}{ll}
p'(a|x),&\text{if $1\leq a\leq m$ and $1\leq x\leq n$} \\ 
\delta(a-m,x-n),&\text{otherwise}\end{array}\right.\;,\\
q(b|y) &=& \left\{\begin{array}{ll}
p'(b|y-n),&\text{if $1\leq b\leq m$ and $n+1\leq y\leq 2n$} \\ 
\delta(b-m,y),&\text{otherwise}\end{array}\right.\;.
\end{eqnarray} 
Roger may now always discern which half of the message set ($\{1,\dots,n\}$ or $\{n+1,\dots,2n\}$) 
that $x$ and $y$ belong to, by simply noting which subdivision $a$ and $b$ belong to 
($\{1,\dots,m\}$ or $\{m+1,\dots,m+n\}$), and immediately calls 0 when different, or performs $r'$ appropriately 
when the same. The resulting worst-case error probability is then equal to that for the $(n,m,n)$-strategy $(p',r')$. 
In Table~\ref{Pwcemotable2} the upper bound $\Pwcemo(10,9,9)\leq\Pwcemo(5,4,5)=1/2$ comes from this method. 

To obtain further upper bounds we can consider a variation of the constant-weight-code strategies given 
in Sec.~\ref{specsec}: If $\Ia^{(p)}=\big\{\Is_x^{(p)}\big\}$ and $\Ia^{(q)}=\big\{\Is_y^{(q)}\big\}$ are 
ordered families of, respectively, $k_1$- and $k_2$-subsets of the $m$-set, both of size $n$, with the property 
that 
\begin{equation}
\Big|\Big(\Is_x^{(p)}\times\Is_y^{(q)}\Big)\cap\Big(\cup_z\:\Is_z^{(p)}\times\Is_z^{(q)}\Big)\Big|\;\leq\; j
\end{equation} 
for all $x\neq y$, then $\Pwcemo(n,m,m)\leq j/(k_1k_2)$. The strategies which achieve this worst-case error 
probability are defined by simply setting $p(a|x)=1/k_1$ when $a\in\Is_x^{(p)}$ (zero otherwise), 
and $q(b|y)=1/k_2$ when $b\in\Is_y^{(q)}$. Let us denote by $N_2(m,k_1,k_2,j)$ the largest $n$ such that 
a pair of such families exist. The following theorem is then a variation of Theorem~\ref{constantweightbound}. 
\begin{thm}
$\Pwcemo(n,m,m)\leq j/(k_1k_2)\;$ if $\;n\leq N_2(m,k_1,k_2,j)\;$.
\end{thm}
Our own computational searches have found that $N_2(5,2,2,3)=8$. A strategy with the corresponding worst-case 
error probability of $3/4$ is defined through the pair of ordered families 
with incidence matrices
\begin{equation}
\Is_x^{(p)}:\quad\begin{array}{c|ccccc}
x \,&\, 1 & 2 & 3 & 4 & 5 \\\hline
1 \,&\, 1 & 0 & 1 & 0 & 0 \\
2 \,&\, 1 & 0 & 0 & 1 & 0 \\
3 \,&\, 1 & 0 & 0 & 0 & 1 \\
4 \,&\, 0 & 1 & 1 & 0 & 0 \\
5 \,&\, 0 & 1 & 0 & 1 & 0 \\
6 \,&\, 0 & 1 & 0 & 0 & 1 \\
7 \,&\, 1 & 1 & 0 & 0 & 0 \\
8 \,&\, 0 & 0 & 1 & 1 & 0
\end{array}\;,\qquad\qquad 
\Is_y^{(q)}:\quad\begin{array}{c|ccccc}
y \,&\, 1 & 2 & 3 & 4 & 5 \\\hline
1 \,&\, 1 & 0 & 1 & 0 & 0 \\
2 \,&\, 1 & 0 & 0 & 1 & 0 \\
3 \,&\, 1 & 0 & 0 & 0 & 1 \\
4 \,&\, 0 & 1 & 1 & 0 & 0 \\
5 \,&\, 0 & 1 & 0 & 1 & 0 \\
6 \,&\, 0 & 1 & 0 & 0 & 1 \\
7 \,&\, 0 & 0 & 1 & 1 & 0 \\ 
8 \,&\, 1 & 1 & 0 & 0 & 0
\end{array}\;.
\label{gen1}\end{equation}
Further computational searches have revealed that $N_2(6,2,2,3)=12$, $N_2(6,2,3,5)=13$, 
$N_2(7,2,2,3)=21$ and $N_2(7,3,3,8)\geq 26$. 
The remaining upper bounds given in Table~\ref{Pwcemotable2} follow from these results.

Finally, returning to the general case, we remark that the lower bounds in Table~\ref{Pwcemotable}
and upper bounds in Table~\ref{Pwcemotable2}, give respectively, lower and upper bounds on 
$\Pwcemo(n,m_A,m_B)$, if we identify $m=\min\{m_A,m_B\}$ in both tables. This follows from the 
trivial inequality $\Pwcemo(n,m_A,m_B-1)\geq\Pwcemo(n,m_A,m_B)\geq\Pwcemo(n,m_A+1,m_B)$. 
The inequalities $\Pwcemo(n+1,m_A,m_B)\geq\Pwcemo(n,m_A,m_B)\geq\Pwcemo(n+1,m_A+1,m_B+1)$  
also hold in the general case.

\section{Quantum fingerprinting}
\label{quantumsec}

\subsection{One-way model: $m_B=n$}

In the quantum fingerprinting scenario, Alice's and Bob's classical fingerprints ($a$ and $b$) and 
probability distributions [$p(a|x)$ and $q(b|y)$] are replaced by quantum states $\rho(x)\in\Q{m_A}$ and $\tau(y)\in\Q{m_B}$, respectively, where 
$\Q{d}\equiv\left\{\rho\in\operatorname{End}(\C^d)\,|\,\rho\geq 0\,,\,\tr(\rho)=1\right\}$. Quantum fingerprinting becomes 
classical when $\rho(x)$ and $\tau(y)$ are diagonal in the computational basis, in which case 
we may identify $p(a|x)=\bra{a}\rho(x)\ket{a}$ and $q(b|y)=\bra{b}\tau(y)\ket{b}$. 

We will first investigate quantum fingerprinting when $m_B=n$, in which case Bob encodes Sapna's messages into orthogonal 
quantum states, upon which Roger may perform an orthogonal measurement to unambiguously determine the state, 
and thus $y$. Though implemented with quantum states, this portion of the fingerprinting procedure is 
effectively classical. Alice, however, must choose her fingerprints from a Hilbert space of 
dimension $m_A=m<n$. Roger's task is then to test whether the state received from Alice, $\rho(x)$, is 
$\rho(y)$ or not, with one-sided error, and thus the projective measurement $\{P(y),1-P(y)\}$, where 
$P(y)$ projects onto the support of $\rho(y)$, is the only suitable choice. The worst-case error probability 
for this strategy is  
\begin{equation}
\Pwce = \max_{x\neq y} \,\tr \left[\rho(x)P(y)\right]\;. \label{PwceOW}
\end{equation}

It is intuitively reasonable that Alice should fingerprint with pure states, and thus, we will first consider 
this case. The worst-case error probability is then simply the maximal pairwise overlap for 
Alice's choice of fingerprint states $\{\ket{\psi(x)}\}_{x=1}^n\subset\C^m$ :
\begin{equation}
\Pwce = \max_{x\neq y} \,|\braket{\psi(x)}{\psi(y)}|^2\;.
\end{equation}
Consequently, the minimum possible worst-case error probability will be achieved when Alice chooses her set of
fingerprint states to have the minimum possible maximal pairwise overlap. Defining the quantity
\begin{equation}
\delta^2(n,m)\;\equiv\;\min_{\{\ket{\psi_k}\}_{k=1}^n\subset\C^m}\;\max_{j\neq k}\; |\braket{\psi_j}{\psi_k}|^2 \;,
\end{equation}
our goal is to now find sets of $n$ pure quantum states in $\C^m$ which achieve the minimum, called 
optimal {\it Grassmannian packings} \cite{conway,barg,strohmer}. We will assume that $n\geq m>1$ in the following. 

Lower bounds on $\delta^2(n,m)$ follow from an isometric embedding of the space of quantum states into 
Euclidean space \cite{conway,barg}. Each $\rho$ is first associated with a traceless Hermitian matrix under the 
correspondence $A(\rho)\equiv\sqrt{\frac{m}{m-1}}\left(\rho-\frac{1}{m}I\right)$. Endowed 
with the Frobenius norm $\|A\|_F\equiv\sqrt{\tr(A^\dag A)}$, the set of all traceless Hermitian matrices 
forms a real normed vector space of dimension $m^2-1$, in which the images of pure states lie on the unit sphere 
$\|A\|_F=1$. Through an appropriate parametrization of $A$, the Frobenius distance for traceless 
Hermitian matrices corresponds to Euclidean distance in $\R^{m^2-1}$. For example, taking 
$x_k(A)=\tr(\lambda_kA)$, where the lambda matrices $\lambda_k$ ($k=1,\dots,m^2-1$) form a 
traceless Hermitian operator basis (see Appendix A of \cite{rungta}), we find that 
$\|A-B\|_F=\|\bm{x}(A)-\bm{x}(B)\|_2$. Consequently, given that for pure states 
$\|A(\psi)-A(\phi)\|_F^2=\frac{2m}{m-1}(1-|\braket{\psi}{\phi}|^2)$, a bound on the maximum possible minimal 
distance between $n$ points on the unit sphere in $\R^{m^2-1}$ will imply a bound on $\delta^2(n,m)$. 

Packings of points on the unit sphere in $\R^d$ are called {\it spherical codes}~\cite{conway2}. The Rankin 
bounds for spherical codes~\cite{rankin} imply for Grassmannian packings, the {\it simplex bound}~\cite{conway,barg}
\begin{equation}
\delta^2(n,m) \geq \frac{n-m}{m(n-1)}\;,
\label{simplexbound}\end{equation}
with equality possible only when $n\leq m^2$, and the {\it orthoplex bound}
\begin{equation}
\delta^2(n,m) \geq \frac{1}{m}\;,
\label{orthoplexbound}\end{equation}
if $n>m^2$, with equality possible only when $n\leq 2(m^2-1)$.

\begin{table}
{\small
\begin{tabular}{|c|@{\hspace{0.01cm}}c@{\hspace{0.01cm}}c@{\hspace{0.01cm}}c@{\hspace{0.01cm}}c@{\hspace{0.01cm}}c@{\hspace{0.01cm}}c@{\hspace{0.01cm}}c@{\hspace{0.01cm}}c@{\hspace{0.01cm}}c@{\hspace{0.01cm}}c@{\hspace{0.01cm}}c@{\hspace{0.01cm}}c@{\hspace{0.01cm}}c@{\hspace{0.01cm}}c@{\hspace{0.01cm}}c@{\hspace{0.01cm}}|}\hline
  $n\backslash m$ & 2           & 3                           & 4                           & 5                           & 6                           & 7                           & 8                           & 9                           & 10                           & 11                           & 12                           & 13                           & 14                           & 15 & 16 \\ \hline 
 2 & 0                          & 0                           & 0                           & 0                           & 0                           & 0                           & 0                           & 0                           & 0                            & 0                            & 0                            & 0                            & 0                            & 0 & 0 \\
 3 & $\phantom{{}^e}\bm{1/4}^e$ & 0                           & 0                           & 0                           & 0                           & 0                           & 0                           & 0                           & 0                            & 0                            & 0                            & 0                            & 0                            & 0 & 0 \\
 4 & $\phantom{{}^e}\bm{1/3}^e$ & $\phantom{{}^e}\bm{1/9}^e$  & 0                           & 0                           & 0                           & 0                           & 0                           & 0                           & 0                            & 0                            & 0                            & 0                            & 0                            & 0 & 0 \\ 
 5 & $\bm{1/2}$                 & $.1886$                & $\phantom{{}^e}\bm{1/16}^e$ & 0                           & 0                           & 0                           & 0                           & 0                           & 0                            & 0                            & 0                            & 0                            & 0                            & 0 & 0 \\
 6 & $\phantom{{}^m}\bm{1/2}^m$ & $\phantom{{}^e}\bm{1/5}^e$  & $.1071$                & $\phantom{{}^e}\bm{1/25}^e$ & 0                           & 0                           & 0                           & 0                           & 0                            & 0                            & 0                            & 0                            & 0                            & 0 & 0 \\
 7 & $\bm{.6051}$          & $\phantom{{}^e}\bm{2/9}^e$  & $\phantom{{}^e}\bm{1/8}^e$  & $.0709$                & $\phantom{{}^e}\bm{1/36}^e$ & 0                           & 0                           & 0                           & 0                            & 0                            & 0                            & 0                            & 0                            & 0 & 0 \\
 8 & $\bm{.6306}$          & $1/4$                       & $\phantom{{}^e}\bm{1/7}^e$  & $.0871$                & $.0502$                & $\phantom{{}^e}\bm{1/49}^e$ & 0                           & 0                           & 0                            & 0                            & 0                            & 0                            & 0                            & 0 & 0 \\
 9 & $\bm{2/3}$                 & $\phantom{{}^e}\bm{1/4}^e$  & $.1615$                & $.1025$                & $\phantom{{}^e}\bm{1/16}^e$ & $.0384$                & $\phantom{{}^e}\bm{1/64}^e$ & 0                           & 0                            & 0                            & 0                            & 0                            & 0                            & 0 & 0 \\
10 & $\bm{.7022}$          & $\bm{1/3}$                  & $.1687$                & $\phantom{{}^e}\bm{1/9}^e$  & $.0741$                & $.0491$                & $.0299$                & $\phantom{{}^e}\bm{1/81}^e$ & 0                            & 0                            & 0                            & 0                            & 0                            & 0 & 0 \\
11 & $\bm{.7236}$          & $\bm{1/3}$                  & $.1808$                & $\phantom{{}^e}\bm{3/25}^e$ & $\phantom{{}^e}\bm{1/12}^e$ & $.0573$                & $.0394$                & $.0238$                & $\phantom{{}^e}\bm{1/100}^e$ & 0                            & 0                            & 0                            & 0                            & 0 & 0 \\
12 & $\bm{.7236}$          & $\phantom{{}^m}\bm{1/3}^m$  & $.1830$                & $.1277$                & $\phantom{{}^e}\bm{1/11}^e$ & $.0651$                & $.0456$                & $.0319$                & $.0193$                 & $\phantom{{}^e}\bm{1/121}^e$ & 0                            & 0                            & 0                            & 0 & 0 \\
13 & $\bm{.7713}$          & $.3871$                & $\phantom{{}^e}\bm{3/16}^e$ & $.1351$                & $.0972$                & $.0714$                & $.0522$                & $\phantom{{}^e}\bm{1/27}^e$ & $.0265$                 & $.0163$                 & $\phantom{{}^e}\bm{1/144}^e$ & 0                            & 0                            & 0 & 0 \\
14 & $\bm{.7820}$          & $.4066$                & $1/5$                       & $.1411$                & $.1026$                & $\phantom{{}^e}\bm{1/13}^e$ & $.0577$                & $.0429$                & $.0310$                 & $.0222$                 & $.0138$                 & $\phantom{{}^e}\bm{1/169}^e$ & 0                            & 0 & 0 \\
15 & $.7963$               & $.4141$                & $1/5$                       & $.1452$                & $.1072$                & $\phantom{{}^e}\bm{4/49}^e$ & $\phantom{{}^e}\bm{1/16}^e$ & $.0476$                & $.0357$                 & $.0262$                 & $.0189$                 & $.0118$                 & $\phantom{{}^e}\bm{1/196}^e$ & 0 & 0 \\
16 & $.8061$               & $.4196$                & $\phantom{{}^e}\bm{1/5}^e$  & $.1506$                & $\phantom{{}^e}\bm{1/9}^e$  & $.0857$                & $\phantom{{}^e}\bm{1/15}^e$ & $.0519$                & $\phantom{{}^e}\bm{1/25}^e$  & $.0303$                 & $\phantom{{}^e}\bm{1/45}^e$  & $.0162$                 & $.0101$                 & $\phantom{{}^e}\bm{1/225}^e$ & 0 \\
17 & $.8140$               & $.4282$                & $\bm{1/4}$                  & $.1549$                & $.1154$                & $.0893$                & $.0703$                & $.0556$                & $.0438$                 & $.0341$                 & $.0261$                 & $.0196$                 & $.0141$                 & $.0089$ & $\phantom{{}^e}\bm{1/256}^e$ \\
18 & $.8243$               & $.4395$                & $\bm{1/4}$                  & $.1581$                & $.1194$                & $.0926$                & $.0735$                & $\phantom{{}^e}\bm{1/17}^e$ & $.0471$                 & $.0374$                 & $.0294$                 & $.0227$                 & $.0173$                 & $.0123$ & $.0079$ \\
19 & $.8366$               & $.4576$                & $\bm{1/4}$                  & $.1581$                & $.1220$                & $.0954$                & $.0764$                & $\phantom{{}^e}\bm{5/81}^e$ & $\phantom{{}^e}\bm{1/20}^e$  & $.0404$                 & $.0324$                 & $.0257$                 & $.0199$                 & $.0154$ & $.0111$ \\
20 & $.8382$               & $.4712$                & $\phantom{{}^m}\bm{1/4}^m$  & $.1581$                & $.1254$                & $.0983$                & $.0790$                & $.0643$                & $\phantom{{}^e}\bm{1/19}^e$  & $.0431$                 & $.0351$                 & $.0284$                 & $.0226$                 & $.0176$ & $.0137$ \\
21 & $.8497$               & $.4733$                & $.2890$                & $\phantom{{}^e}\bm{4/25}^e$ & $.1271$                & $.1011$                & $.0813$                & $.0667$                & $.0550$                 & $.0455$                 & $.0375$                 & $.0308$                 & $.0250$                 & $.0200$ & $\phantom{{}^e}\bm{1/64}^e$ \\
22 & $.8552$               & $.4948$                & $.2979$                & $1/6$                       & $.1293$                & $.1039$                & $.0835$                & $.0688$                & $.0571$                 & $.0476$                 & $.0397$                 & $.0330$                 & $.0272$                 & $.0222$ & $.0179$ \\\hline
\end{tabular}}
\caption{The smallest known maximal pairwise overlap for sets of $n$ pure states in $\C^m$. Entries in 
boldface signify that the overlap is optimal. This is the case for all ETFs (denoted by a superscript $e$) 
and MUBs ($m$). Each maximal overlap is also an achievable worst-case error probability for a quantum 
fingerprinting strategy under the one-way communication model ($m_B=n$), with the entries in boldface being 
the minimum achievable.}
\label{linepacktable}\end{table}

The simplex bound was derived by Welch~\cite{welch} and also appears in the context of frame theory 
\cite{strohmer,holmes} where a set of states which achieves equality is called an 
{\it equiangular tight frame} (ETF). This name derives from the fact that equality will occur if 
and only if~\cite{strohmer}
\begin{equation}
|\braket{\psi_j}{\psi_k}|^2 = \frac{n-m}{m(n-1)}\;,
\label{ETF}\end{equation}
for all $j\neq k$, which automatically guarantees a tight frame: $\sum_k \ket{\psi_k}\bra{\psi_k} = \frac{n}{m}I$. 
Trivial examples of ETFs occur when $n=m$ (basis in $\C^m$) and $n=m+1$ (vertices of the simplex in $\R^m$).
Excluding these cases, it is necessary that $n\leq \min \{m^2,(n-m)^2\}$ for an ETF to exist \cite{strohmer,holmes}. 
This condition, however, is far from sufficient. In Table~\ref{linepacktable} we give all known examples of an 
ETF for small $m$ and $n$. Here we tabulate upper bounds on $\delta^2(n,m)$. The entries in Table~\ref{linepacktable} 
marked by a superscript $e$ signify cases where an analytical construction of an ETF is known. These derive 
from quadratic residues \cite{zauner,renes}, conference matrices \cite{delsarte,strohmer,holmes,bodmann,renes}, 
Hadamard matrices \cite{holmes,bodmann}, graphs \cite{delsarte,holmes,sustik}, and difference 
sets \cite{xia,konig}. Finally, an ETF is conjectured to exist whenever $n=m^2$ \cite{zauner,renes2}, with 
analytical constructions for $m\leq 8$ and $m=19$ \cite{zauner,renes2,hoggar,grassl,appleby}.

Given the simplex bound on $\delta^2(n,m)$ [Eq.~(\ref{simplexbound})], we know that all ETFs, when they 
exist, are optimal Grassmannian packings. Sets of mutually unbiased bases (MUBs) \cite{wootters} also give 
optimal Grassmannian packings. It is known that when $m$ is a prime power, a maximal set of $m+1$ MUBs exists 
(i.e. $n=m^2+m$) with constant overlap of $1/m$ between elements of different bases, thus saturating 
the orthoplex bound [Eq.~\ref{orthoplexbound}]. In Table~\ref{linepacktable} the MUBs are marked by a 
superscript $m$.

When $m=2$, the above embedding maps quantum states surjectively into the unit ball in $\R^3$ (realizing the Bloch-sphere representation of a qubit), 
and our problem is exactly equivalent to finding optimal spherical codes in three dimensions. Collections of 
putatively optimal spherical codes can be found online~\cite{sloane}, and have the maximal overlaps in the first 
column of Table~\ref{linepacktable}. Of these, only in the cases $n\leq 14$ has optimality 
been established (see \cite{kottwitz,buddenhagen}). For $n=4$, 6, 8, and 12, the optimal spherical codes 
are given by the vertices of a regular tetrahedron, regular octahedron, square antiprism, and regular icosahedron, 
with maximal overlaps $\delta^2=1/3$, $1/2$, $(3+\sqrt{2})/7$, and $(1+1/\sqrt{5})/2$, respectively \cite{conway2}. 
Except for $n=8$, optimality follows from the Fejes T\'{o}th bound~\cite{toth}: $\delta^2(n,2)\geq \frac{1}{4}\csc^2\big[\frac{\pi n}{6(n-2)}\big]$. 

All other entries in Table~\ref{linepacktable} are the overlaps for the best packings found in a numerical search, 
and thus, optimality is not guaranteed. In many cases, however, the given 
overlap almost saturates the simplex bound and thus must be at least very close to optimal.

When $m_B=n$, each of the maximal overlaps in Table~\ref{linepacktable} is the worst-case error probability for a quantum 
fingerprinting strategy. Comparing these to the classical error probabilities in Table~\ref{Pwcemotable}, we see that quantum 
fingerprinting strategies generally outperform classical strategies by a sizeable margin.
Note, for example, that when $n=q(q+1)$, where $q$ is a prime power, there exists a classical constant-weight-code strategy 
[see Eq.~(\ref{classicalMUBs})] which requires $m=q^2$ fingerprints to obtain a worst-case error probability of $\Pwce=1/q$. 
This class of classical strategies is optimal for at least $q=2$ and $3$. There exist ETFs, however, with the parameters 
$m=q^2$ and $n=q(q+1)+1$ (when $q$ is a prime power) having equiangular overlaps of $1/q^3$ \cite{konig}. Consequently, 
for the same $m$ and $n$, the classical worst-case error probability above may be reduced to $\Pwce=1/q^3$, or less, using quantum 
fingerprints. Taking a different approach, if we instead fix $n=q(q+1)$ and $\Pwce=1/q$, we know that a quantum MUBs strategy 
exists which achieves this performance using only $m=q$ fingerprints. Finally, the quantum performance gain is far more impressive 
if we instead fix $m$ and $\Pwce$. For example, in the classical case, when $m=4$ we know that to achieve $\Pwce=1/2$ it is 
necessary that $n\leq 6$. A numerical search, however, finds that there exists a set of $n$ 4-dimensional quantum states with 
a maximum pairwise overlap of $1/2$, or less, for all $n\leq 75$. Thus, quantum fingerprinting strategies 
can map at least 75 messages into 4 fingerprints while retaining a worst-case error probability $\Pwce\leq 1/2$, compared 
to only 6 messages classically. 

In the above analysis we have assumed that Alice's fingerprint states are pure. Under this restriction, each 
boldfaced error probability in Table~\ref{linepacktable} is the minimum possible. Suppose now that Alice is allowed 
to choose mixed quantum states for her fingerprints. When $m=2$, it is clear from Eq.~(\ref{PwceOW}) 
that pure states remain the only valid choice if she is to retain $\Pwce<1$, since any mixed qubit has support on the entire Hilbert space. 
Consequently, each boldfaced entry in the first column of Table~\ref{linepacktable} is the minimum achievable worst-case 
error probability over {\it all} (mixed or pure) quantum fingerprinting strategies. We will conclude this subsection by proving 
that the remaining boldfaced error probabilities in Table~\ref{linepacktable} are also the minimum achievable. 

Define the quantity 
\begin{equation}
\Delta^2(n,m)\;\equiv\;\min_{\{\rho_k\}_{k=1}^n\subset\Q{m}}\;\max_{j\neq k}\; \,\tr \left(\rho_j\rho_k\right) \;.
\end{equation}
By considering the choice $\rho_k=I/m$ for all $k$, 
we immediately arrive at the upper bound $\Delta^2(n,m)\leq 1/m$. We will now show that the Rankin lower bounds on the maximum pairwise 
overlap for a set of pure quantum states [Eq.'s~(\ref{simplexbound}) and (\ref{orthoplexbound})] also apply to general quantum states. 
First note that
\begin{equation}
\tr \left(\rho_j\rho_k\right)\,=\, \frac{1}{m} + \frac{m-1}{m}\tr \left[A(\rho_j)A(\rho_k)\right] 
\label{someequation}\end{equation}
under the correspondence $A(\rho)\equiv\sqrt{\frac{m}{m-1}}\left(\rho-\frac{1}{m}I\right)$. In the context of deriving 
lower bounds on $\Delta^2(n,m)$, we need only consider sets of quantum states with  
$\tr\left[A(\rho_j)A(\rho_k)\right]\leq 0$ for all $j\neq k$, since otherwise the maximum pairwise overlap is greater than $1/m$, 
which exceeds the upper bound. Assuming $\tr\left[A(\rho_j)A(\rho_k)\right]\leq 0$, and given that $\|A(\rho)\|_F\leq 1$ for any 
quantum state, we see that 
\begin{equation}
\tr\left[A(\rho_j)A(\rho_k)\right] \,\geq\, \tr \big[\tilde{A}(\rho_j)\tilde{A}(\rho_k)\big] \,=\, 1-\frac{1}{2}\|\tilde{A}(\rho_j)-\tilde{A}(\rho_k)\|_F^2\;,
\end{equation}
where the rescaled $\tilde{A}(\rho_k)\equiv A(\rho_k)/\|A(\rho_k)\|_F$ satisfy $\|\tilde{A}(\rho_k)\|_F=1$. Thus, under the above 
isometric embedding into $(m^2-1)$-dimensional Euclidean space, the Rankin bounds on spherical codes~\cite{rankin} imply that
\begin{equation}
\Delta^2(n,m) \geq \frac{n-m}{m(n-1)}\;,
\end{equation}
and 
\begin{equation}
\Delta^2(n,m) = \frac{1}{m}\;,
\end{equation}
if $n>m^2$. Equality in the latter follows from the upper bound. 

Finally, denoting Alice's fingerprint states by $\rho(x)$, and the support of $\rho(x)$ by $P(x)$, the worst-case error probability   
\begin{eqnarray}
\Pwce &=& \max_{x\neq y} \,\tr \left[\rho(x)P(y)\right] \\ 
&\geq& \max_{x\neq y} \,\tr \left[\rho(x)\rho(y)\right] \\
&\geq& \Delta^2(n,m)\;,
\end{eqnarray}
and consequently, each boldfaced error probability in Table~\ref{linepacktable} is in fact the minimum achievable worst-case 
error probability over {\it all} (mixed or pure) quantum fingerprinting strategies.

\subsection{Generalizations and the SMP model: $m_A=m_B$}

Now consider quantum strategies under the general fingerprinting scenario ($m_A,m_B\leq n$). If Alice sends state $\rho(x)$ for message $x$, and 
Bob sends $\tau(y)$, then, given the one-sided-error constraint, it is straightforward to see that Roger's optimal 
measurement on the product state $\rho(x)\otimes\tau(y)$ to determine whether $x=y$, is the projective measurement 
$\{P_1,P_0=1-P_1\}$, where $P_1$ projects onto the support of $\tilde{P}_1\equiv\sum_z\rho(z)\otimes\tau(z)$. The 
worst-case error probability is then 
\begin{equation}
\Pwce = \max_{x\neq y} \,\tr \left[\rho(x)\otimes\tau(y) P_1\right]\;.
\end{equation}
It remains to find good choices for $\rho(x)$ and $\tau(y)$. For simplicity, we will  
restrict our study to strategies where the fingerprint states are pure, $\rho(x)=\ket{\psi(x)}\bra{\psi(x)}$ and 
$\tau(y)=\ket{\phi(y)}\bra{\phi(y)}$ say.  

Consider the case where $m_A=m_B=m$, and suppose Alice and Bob map messages onto the same set of 
fingerprint states, $\{\ket{\psi(x)}\}_{x=1}^n\subset\C^m$ say. If Roger, instead of choosing 
to project onto the support of $\sum_z\ket{\psi(z)}\bra{\psi(z)}\otimes\ket{\psi(z)}\bra{\psi(z)}$, fixes
$P_1=\Pi_{\text{sym}}\equiv\frac{1}{2}\sum_{a,b} \ket{a}\bra{a}\otimes\ket{b}\bra{b}+\ket{a}\bra{b}\otimes\ket{b}\bra{a}$,
the projector onto the symmetric subspace of $\C^m\otimes\C^m$, then the resulting worst-case error probability is
\begin{equation}
\Pwce = \max_{x\neq y} \,\frac{1}{2}\left(1+|\braket{\psi(x)}{\psi(y)}|^2\right) \;.
\label{swaptestPwce}\end{equation}
The optimal Grassmannian packings described in the previous subsection are now the best choices for $\{\ket{\psi(x)}\}_{x=1}^n$. Although 
the minimum possible worst-case error probability is bounded below by one half, $\Pwce=\frac{1}{2}[1+\delta^2(n,m)]\geq 1/2$, 
these strategies were found to give the best error rates when $n\gtrsim\frac{1}{2}m(m+1)$, the dimension of the symmetric subspace.

We now describe analytical constructions of quantum fingerprinting strategies from ETFs under the SMP model. For an ETF, the strategy 
just described has $\Pwce=\frac{1}{2}\big[1+\frac{n-m}{m(n-1)}\big]$. However when $n<\frac{1}{2}m(m+1)$ we can obtain smaller 
worst-case error probabilities in the following way. Suppose the set $\{\ket{\xi_j}\}_{j=1}^n$ forms an ETF in $\C^m$. We choose 
$\ket{\psi(x)}=\ket{\xi_x}$ and $\ket{\phi(y)}=\ket{{\xi_y}^*}$, for Alice's and Bob's fingerprint states respectively, where 
conjugation is done in the computational basis. Note that $\tilde{P}_1\equiv\sum_z\ket{\psi(z)}\bra{\psi(z)}\otimes\ket{\phi(z)}\bra{\phi(z)}$ is 
the sum of $n$ rank-1 operators, and hence, has rank no greater than $n$.
Using the defining property of an ETF [Eq.~(\ref{ETF})], it is straightforward to verify that the $n$ states
\begin{eqnarray}
\ket{\Xi_k}&\equiv&\sqrt{\frac{m(n-1)}{n^2(m-1)}}\sum_{j=1}^ne^{\frac{2\pi ijk}{n}}\ket{\xi_j}\otimes\ket{{\xi_j}^*}\,, \qquad k=1,\dots,n-1\,, \\
\ket{\Xi_n}&\equiv&\sqrt{\frac{m}{n^2}}\sum_{j=1}^n\ket{\xi_j}\otimes\ket{{\xi_j}^*}\,,
\end{eqnarray}
are orthonormal and belong to the support of $\tilde{P}_1$. Consequently, these states {\em span} the support of $\tilde{P}_1$, 
and Roger's optimal measurement has $P_1=\sum_{k=1}^n\ket{\Xi_k}\bra{\Xi_k}$. The worst-case error probability for the strategy may now be calculated in a 
straightforward manner. The result is $\Pwce=\frac{n^2-m^2}{m^2(n-1)}$.

When $n=2m$, ETFs can be used to obtain even smaller worst-case error probabilities. Suppose again that the set $\{\ket{\xi_j}\}_{j=1}^n$ 
forms an ETF in $\C^m$. Define the $m\times n$ matrix with entries $X_{jk}\equiv\sqrt{\frac{m}{n}}\braket{j}{\xi_k}$. The tight frame condition, 
$\sum_k \ket{\xi_k}\bra{\xi_k} = \frac{n}{m}I$, implies that the rows of $X$ are orthonormal: $XX^\dag=I$. Thus, another
$n-m$ rows may be appended to form an $n\times n$ unitary matrix $U=\left[\frac{X}{Y}\right]$, where $XY^\dag=YX^\dag=0$ and $YY^\dag=I$. 
This last relation implies that a tight frame in $\C^{n-m}$ may be constructed from the columns of $Y$ by setting 
$\braket{j}{\chi_k}=\sqrt{\frac{n}{n-m}}Y_{jk}$. The set $\{\ket{\chi_j}\}_{j=1}^n$ is in fact an ETF. This follows from the unitarity 
of $U$: $I=U^\dag U=X^\dag X+Y^\dag Y$, i.e. $\delta_{jk}=\frac{m}{n}\braket{\xi_j}{\xi_k}+\frac{n-m}{n}\braket{\chi_j}{\chi_k}$. 
Thus, when $j\neq k$ we have $|\braket{\chi_j}{\chi_k}|^2=\frac{m^2}{(n-m)^2}|\braket{\xi_j}{\xi_k}|^2=\frac{m}{(n-m)(n-1)}$, which 
is the defining property of an ETF in $\C^{n-m}$. We have shown that the existence of an ETF in $\C^m$ guarantees the 
existence of an ETF (of equal size) in $\C^{n-m}$ \cite{holmes}. Note that for the special case $n=2m$, the second ETF also belongs
to $\C^m$. Importantly, again from the unitarity of $U$, the inner products of the ETF pair satisfy 
$\braket{\chi_j}{\chi_k}=-\braket{\xi_j}{\xi_k}$ for $j\neq k$. Using this relation and Eq.~(\ref{ETF}), it is straightforward to verify that the 
$n-1$ states
\begin{equation}
\ket{\Xi_k}\;\equiv\;\frac{\sqrt{n-1}}{n}\sum_{j=1}^ne^{\frac{2\pi ijk}{n}}\ket{\xi_j}\otimes\ket{{\chi_j}^*}\,, \qquad k=1,\dots,n-1\,, 
\end{equation}
are the orthonormal and belong to the support of $\tilde{P}_1$, where we are choosing $\ket{\psi(x)}=\ket{\xi_x}$ and $\ket{\phi(y)}=\ket{{\chi_y}^*}$ 
for Alice's and Bob's fingerprint states respectively. Unlike in the previous case, $\sum_{j}\ket{\xi_j}\otimes\ket{{\chi_j}^*}=0$, 
which means $\rank\tilde{P}_1\leq n-1$. Consequently, Roger's optimal measurement has $P_1=\sum_{k=1}^{n-1}\ket{\Xi_k}\bra{\Xi_k}$.
The worst-case error probability for this quantum fingerprinting strategy is $\Pwce=\frac{3m-2}{m(2m-1)}$.

\begin{table}
{\small
\begin{tabular}{|c|@{\hspace{0.3cm}}c@{\hspace{0.3cm}}c@{\hspace{0.3cm}}c@{\hspace{0.3cm}}|}\hline
  $n\backslash m$ & 2      & 3                       & 4            \\\hline 
 2 & 0                     & 0                       & 0            \\
 3 & $\phantom{{}^e}5/8^e$ & 0                       & 0            \\
 4 & $\phantom{{}^e}2/3^e$ & $\phantom{{}^e}7/27^e$  & 0            \\ 
 5 & $3/4$                 & $.4330$            & $\phantom{{}^e}9/64^e$         \\
 6 & $\phantom{{}^m}3/4^m$ & $\phantom{{}^e}7/15^e$  & $.2494$ \\
 7 & $.8025$          & $\phantom{{}^e}11/18^e$ & $.3398$ \\
 8 & $.8153$          & $5/8$                   & $\phantom{{}^e}5/14^e$         \\
 9 & $5/6$                 & $\phantom{{}^e}5/8^e$   & $.4962$ \\
10 & $.8511$          & $2/3$                   & $.5567$ \\
11 & $.8618$          & $2/3$                   & $.5904$ \\
12 & $.8618$          & $\phantom{{}^m}2/3^m$   & $.5915$ \\
13 & $.8857$          & $.6935$            & $\phantom{{}^e}19/32^e$        \\
14 & $.8910$          & $.7033$            & $3/5$          \\
15 & $.8982$          & $.7071$            & $3/5$          \\
16 & $.9031$          & $.7098$            & $\phantom{{}^e}3/5^e$          \\
17 & $.9070$          & $.7141$            & $5/8$          \\
18 & $.9122$          & $.7198$            & $5/8$          \\
19 & $.9183$          & $.7288$            & $5/8$          \\
20 & $.9191$          & $.7356$            & $\phantom{{}^m}5/8^m$          \\
21 & $.9249$          & $.7367$            & $.6445$ \\
22 & $.9276$          & $.7474$            & $.6490$ \\\hline
\end{tabular}}
\caption{The smallest known worst-case error probabilities for a quantum fingerprinting strategy 
under the SMP model ($m_A=m_B=m$).}
\label{SMPtable}\end{table}

In Table~\ref{SMPtable}, putatively optimal worst-case error probabilities are collected for $m=2,3$ and $4$. Strategies which 
use ETFs or MUBs are marked by a superscript $e$ or $m$, respectively. When $n<\frac{1}{2}m(m+1)$ some of the analytical quantum
fingerprinting strategies described above are outperformed by those found numerically, and the corresponding numerical worst-case error probability 
is given instead. Numerical strategies were located by setting $P_1$ to the support of 
$\tilde{P}_1\equiv\sum_z\ket{\psi(z)}\bra{\psi(z)}\otimes\ket{\phi(z)}\bra{\phi(z)}$ as the fingerprint sets $\{\ket{\psi(x)}\}_{x=1}^n$ 
and $\{\ket{\phi(y)}\}_{y=1}^n$ were varied across Hilbert space.   
Although none of the quantum error rates in Table~\ref{SMPtable} can be claimed optimal, the lower bounds on the classical 
error rates in Table~\ref{Pwcemotable2} are still surpassed by a fair margin. Again, the performance gain is most impressive 
when we consider $m$ and $\Pwce$ fixed. For example, in the classical case, when $m=4$ we know that to achieve $\Pwce=3/4$ it is 
necessary that $n\leq 6$. There exists a set of $n$ 4-dimensional quantum states, however, with 
a maximum pairwise overlap of $1/2$, or less, for all $n\leq 75$. Thus, quantum fingerprinting strategies 
can map at least 75 messages into 4 fingerprints while retaining $\Pwce\leq 3/4$ 
[see Eq.~(\ref{swaptestPwce})], compared to only 6 messages classically.  

\section{Conclusion}
\label{concludesec}

In this article we have derived lower bounds on the worst-case error probability for a classical fingerprinting protocol
with one-sided error which are applicable in the small-message limit. These are our main results and the content of Theorems~\ref{antibound}, \ref{coverfreebound}, 
\ref{coverfreebound2} and \ref{antibound2}. Although the majority of the lower bounds are tight only in the one-way communication model, 
they also apply to a generalized scenario which encompasses the simultaneous message passing model. Furthermore, the set-theoretic techniques used 
for their derivation might be of interest from an asymptotic point of view. Additionally, we have presented quantum fingerprinting protocols derived from spherical codes, equiangular tight frames and mutually unbiased bases, with error rates 
surpassing the classical bounds. We hope that our work provides some important new results applicable to current experimental 
investigations of quantum fingerprinting protocols \cite{Horn04,Du04}. The absolute limits of successful fingerprinting protocols provide 
quantitative measures for the compressibility of information stored in message strings. Our analysis may be appended to the growing list 
which reveal a fundamentally greater capacity to compress data stored as quantum information.

\begin{acknowledgments}
AJS would like to thank Harry Buhrman, Nicholas Cavenagh, Dmitry Gavinsky and John Watrous for helpful discussions. 
This work has been supported by CIAR, CSE, iCORE and MITACS. JW acknowledges support from AIF and PIMS.
\end{acknowledgments}

\end{document}